\documentclass[twoside]{article}

\makeatletter
\renewcommand{\@oddhead}%
 {\raisebox{0pt}[\headheight][0pt]%
  {\vbox%
   {\hbox to \textwidth{\hfil%
    {\small\it Quantum geometrodynamics in extended phase space -- II}%
    \strut \hfil}%
   \hrule}}}
\renewcommand{\@evenhead}%
 {\raisebox{0pt}[\headheight][0pt]%
  {\vbox%
   {\hbox to \textwidth{\hfil%
    {\small\it V.A. Savchenko, T.P. Shestakova and G.M. Vereshkov}%
    \strut \hfil}%
   \hrule}}}
\renewcommand{\@oddfoot}{\hfil \thepage \hfil}
\renewcommand{\@evenfoot}{\hfil \thepage \hfil}

\newcounter{sect}
\newcommand{\sect}[1]%
 {\par\vspace*{4mm plus 1mm minus .5mm}\refstepcounter{sect}
  \begin{center}
   \large\bf \thesect.\hspace{2pt}#1
   \samepage
  \end{center}}
\renewcommand{\thesect}{\arabic{sect}}

\makeatother

\begin{document}

\thispagestyle{plain}
\renewcommand{\thefootnote}{\fnsymbol{footnote}}

\begin{center}
{\Large\bf QUANTUM GEOMETRODYNAMICS\\[2pt]
           IN EXTENDED PHASE SPACE -- II.\\[10pt]
{\large\sl THE BIANCHI IX MODEL}

\vspace{10mm}
V. A. Savchenko\footnote{e-mail: savchenko@phys.rnd.runnet.ru}
T. P. Shestakova\footnote{e-mail: stp@phys.rnd.runnet.ru}
and G. M. Vereshkov}

\vspace{5mm}
{\it Dept. of Theoretical Physics, Rostov State University,
     Sorge Str. 5, Rostov-on-Don, Russia}
\end{center}

\begin{abstract}
\small The mathematically correct approach to constructing quantum
geometrodynamics of a closed universe formulated in Part I is realized on the
cosmological Bianchi-IX model with scalar fields. The physical adequacy of the
obtained gauge-noninvariant theory to existing concepts about possible
cosmological scenarios is shown. It is demonstrated that the Wheeler-DeWitt
quantum geometrodynamics based on general quantum theoretical principles
with probability interpretation of a closed universe wave function does not
exist. The problem of the creation of the Universe is considered as a
computational problem of a quantum reduction of the singular state ``Nothing"
to one of possible initial physical quantum states.
\end{abstract}

\begin{center}
\small PACS: 04.60.Ds, 04.60.Gw, 04.60.Kz, 98.80.Hw, 98.80.Bp.
\end{center}
\date{}

\setcounter{footnote}{0}
\renewcommand{\thefootnote}{\arabic{footnote}}

\vspace{5mm}
\noindent In Part I of this work \cite{Part_I} a mathematically correct
gauge-noninvariant approach to quantum geometrodynamics (QGD) of a closed
universe has been developed. Operationally it aimed at describing an
integrated system including a physical object and observation means (OM).
Presented below Part II contains detailed mathematical proofs of the
possibility to apply this approach to simple but representative Bianchi-IX
cosmology as a model.

The model, as well as the chosen class of gauge conditions and
parametrizations are discussed in Sec.\,\ref{Model}. The Lagrangian dynamics
is considered in Sec.\,\ref{dynamics} and in the same section the
Hamiltonian dynamics in extended phase space (EPS) is constructed that is
completely equivalent to the Lagrangian one. The possibility to construct
the Hamiltonian dynamics in this approach is ensured by a special choice of
a gauge condition in the differential form.

In Sec.\,\ref{BFV} we construct the Hamiltonian formulation along the
Batalin -- Fradkin -- Vilkovisky (BFV) line and compare the two approaches.
We point out to the main mathematical reason why the two approaches turn out
not to be equivalent: the reason is that the group of gauge transformations
and the group of transformations generated by constraints are different. It
results in a different structure of ghost sectors in the Lagrangian and
Hamiltonian formulations. At the same time, the two approaches are in
agreement in a gauge-invariant sector which can be singled out by asymptotic
boundary conditions. In the case of a close universe one cannot appeal to
the assumption about asymptotic states and the two formulations of the
Hamiltonian dynamics in EPS lead to different results. Our preference is
given to the approach presented in Sec.\,\ref{dynamics}.

In Sec.\,\ref{GVCrole}, \ref{Class.exact} a probable role of the
gravitational vacuum condensate (GVC) in cosmological evolution is discussed
and illustrated for the exact solution of conditionally-classical dynamical
equations.

In Sec.\,\ref{Quant.dyn} the Schr\"odinger equation for a wave function
of the Universe is derived from a path integral by a standard method
originated from Feynman; the Hamiltonian operator in the Schr\"odinger
equation corresponds to the Hamiltonian constructed in
Sec.\,\ref{dynamics}. The structure of the general solution to the
Schr\"odinger equation is analyzed in Sec.\,\ref{GS}.

In Sec.\,\ref{WDWprobl} we discuss the possibility to go over from the
formulation of QGD in extended phase space to the Wheeler -- DeWitt theory.
A special attention is paid to the demand of BRST-invariance of the
wave function. It is shown that in the approach presented here this demand
leads to the Wheeler -- DeWitt equation only if the assumption about
asymptotic states is made. At the same time, the known Wheeler -- DeWitt
theory is demonstrated to be implicitly gauge-noninvariant and undeducible
correctly, as a physical theory of a closed universe, from general quantum
theoretical principles.

The exact solution to the Schr\"odinger equation corresponding to the
conditially-classical exact solution (Sec.\,\ref{Class.exact}) is found
in Sec.\,\ref{SE_exact}.

The verisimilitude of the physical content of the exact solution from the
standpoint of existing concepts about cosmological scenarios justifies
the attempt undertaken in Sec.\,\ref{Creation} to extrapolate the developed
above methodology to the problem of the creation of the Universe.

The two parts of the work have a through section numbering.

\setcounter{sect}{8}
\sect{The Bianchi IX model}
\label{Model}
The interval in the cosmological Bianchi IX model looks like \cite{Lan}
\begin{equation}
\label{ds}
ds^2=N^2(t)\,dt^2-\eta_{ab}(t)e^a_ie^b_kdx^idx^k;
\end{equation}
\begin{equation}
\label{eta_ab}
\eta_{ab}(t)=\mathop{\rm diag}\nolimits\left(a^2(t),b^2(t),c^2(t)\right),
\end{equation}
$$
\begin{array}{l}
e^1_i=(\sin x^3,\;-\cos x^3\sin x^1,\;0),\\
e^2_i=\rule{0pt}{14pt}(\cos x^3,\;\sin x^3\sin x^1,\;0),\\
e^3_i=\rule{0pt}{14pt}(0,\;\cos x^1,\;1).
\end{array}
$$
We shall also assume that the model includes an arbitrary number $K$ of real
scalar fields described by the Lagrangian
$$
L_{(scal)}=
 \frac1{2\pi^2}\sqrt{-g}\,\left[\frac12\sum_{i=1}^K
  \phi^i_{,\mu}\phi_i^{,\mu}-U_s(\phi_1,\ldots,\phi_K)\right],
$$
and use the following parametrization:
$$
\begin{array}{l}
a=\displaystyle\frac12 r\exp\left[
 \frac12\left(\sqrt{3}\,\varphi+\chi\right)\right];\\
b=\rule{0pt}{20pt}\displaystyle\frac12 r\exp\left[
 \frac12\left(-\sqrt{3}\,\varphi+\chi\right)\right];\\
c=\rule{0pt}{20pt}\displaystyle\frac12 r\exp\left(-\chi\right).
\end{array}
$$

Writing out the Einstein equations in the given parametrization it is easy
to notice \cite{Duncan} that the Bianchi IX model can be considered as a
model of a Friedman-Robertson-Walker closed universe with $r(t)$ being a
scale factor, on which a transversal nonlinear gravitational wave $\varphi(t),
\chi(t)$ is superposed.

The action for a system of the scalar fields and gravitation can be
presented in the simple form
$$
S_0=\int dt\left[\frac12\,\mu_0\gamma_{ab}\,\dot{Q}^a\dot{Q}^b
 -\frac1{\mu_0}U(Q)\right],
$$
where
\begin{equation}
\label{U,Ug,Us}
U(Q)=e^{2q}\,U_g(\varphi,\chi)+e^{3q}\,U_s(\phi),
\end{equation}
$$
U_g(\varphi,\chi)=
 \frac23\left\{\exp\left[2\left(\sqrt{3}\,\varphi+\chi\right)\right]
 +\exp\left[2\left(-\sqrt{3}\,\varphi+\chi\right)\right]
 +\exp(-4\chi)-\right.
$$
$$
 \left. -2\exp\left[-\left(\sqrt{3}\,\varphi+\chi\right)\right]
 -2\exp\left(\sqrt{3}\,\varphi-\chi\right)
 -2\exp(2\chi)\right\},
$$
\begin{equation}
\label{mu0}
\mu_0=\frac{r^3}N,
\end{equation}
$$
r=\exp\left(\frac q2\right),
$$
$$
Q^a=(q,\,\varphi,\,\chi,\,\phi,\,\ldots);
$$
indices $a,b,\ldots$ are raised and lowered with the ``metric"
$$
\gamma_{ab}=\mathop{\rm diag}\nolimits(-1,\,1,\,1,\,1,\,\ldots);
$$

A gauge variable should not inevitably be determined by the relation
(\ref{mu0}). Hereinafter we shall adhere to the parametrization
\begin{equation}
\label{mu0,v}
\mu_0=v(\mu,Q).
\end{equation}

In the present work we shall confine our attention to the special class of
gauges not depending on time,
\begin{equation}
\label{mu,f,k}
\mu=f(Q)+k;\quad
k={\rm const},
\end{equation}
or, in a differential form,
\begin{equation}
\label{mudot}
\dot{\mu}=f_{,a}\dot{Q}^a,
\end{equation}
an index after a comma here and further denoting differentiation with respect
to generalized coordinates:
$f_{,a}=\partial f/\partial Q^a$. Eq.\,(\ref{mudot}) is a
model form of the general gauge (10), Sec.\,6, introducing the missing
velocity $\dot{\mu}$ to the Lagrangian and thus enabling to go over to
Hamiltonian dynamics in EPS \cite{Hennaux}.
Practically, any gauge can be represented by Eq.\,(\ref{mudot}) using an
appropriate parametrization (\ref{mu0,v}).

The latter reflects an obvious fact that the choice of a gauge variable
parametrization and the choice of a gauge condition have an inseparable
interpretation: they are both determined by a clock's construction which time
counting depends on, the influence of the physical subsystem with coordinates
$Q^a$ on the clock going being taken into account. In particular, without
loss of generality any gauge condition can be turned to $\mu={\rm const}$ by
choosing the function $v$. In other words, splitting the definition procedure
of a gauge variable into choosing of parametrization and imposing a gauge
condition is quite conventional, and the familiar reparametrization
noninvariance of the Wheeler -- DeWitt equation \cite{DeWitt} (see, for
example, \cite{HP}) is essentially ill-hidden gauge noninvariance (see also
Sec.\,\ref{WDWprobl}).

\sect{The Lagrangian and Hamiltonian dynamics}
\label{dynamics}
The ghost action corresponding to a gauge from the class (\ref{mudot}) reads
$$
S_{ghost}=
 -\!\int\!dt\,i\bar\theta\frac d{dt}\left[w(\mu,Q)\dot\theta
  -\left(\dot\mu-f_{,a}\dot Q^a\right)\theta\right].
$$
here
$w(\mu,Q)\equiv v(\mu,Q)/v_{,\mu};\;
v_{,\mu}=\partial v(\mu,Q)/\partial\mu;\;
\theta,\;\bar{\theta}$
are the Faddeev-Popov ghosts after replacement
$\bar{\theta}\to-i\bar{\theta}$. The convenience of the latter is that
the Lagrangian is real (Hermitian) when the variables $\theta,\bar{\theta}$
are real, on account of the known complex conjugation rule
for Grassmannian variables:
$(\bar{\theta}\theta)^*=\theta^*\bar{\theta}^*.$

It it convenient to write the ghost action in the dynamically-equivalent
form to avoid the appearance of second time derivatives in the Lagrangian.
Then, redefining the Lagrange multiplier $\lambda$ of a gauge-fixing term
we write the effective action in the form
\begin{equation}
\label{Seff}
S_{ef\!f}=
 \!\int\!dt\,\biggl\{\frac12
   v(\mu,Q)\gamma_{ab}\dot Q^a\dot Q^b
  -\frac 1{v(\mu,Q)}U(Q)
  +\pi\left(\dot\mu-f_{,a}\dot Q^a\right)
  +iw(\mu,Q)\dot{\bar\theta}\dot\theta\biggr\},
\end{equation}
where $\pi\equiv\lambda+\dot{\bar\theta}\theta$.

The action (\ref{Seff}), strictly speaking, makes sense in the quantum
theory only. In the path integral approach (\ref{Seff}) yields the extremal
equations
\begin{equation}
\label{Q-eqn}
\left(v(\mu,Q)\dot Q_a\right)^.
 -\frac12v_{,a}\dot Q^b\dot Q_b
 -\frac1{v^2(\mu,Q)}v_{,a}U(Q)
 +\frac1{v(\mu,Q)}U_{,a}
 -\dot\pi f_{,a}
 +iw_{,a}\dot{\bar\theta}\dot\theta=0,
\end{equation}
\begin{equation}
\label{mu-eqn}
\frac12v_{,\mu}\dot Q^a\dot Q_a
 +\frac1{v^2(\mu,Q)}v_{,\mu}U(Q)
 -\dot\pi
 +iw_{,\mu}\dot{\bar\theta}\dot\theta=0,
\end{equation}
\begin{equation}
\label{pi-eqn}
\dot\mu-f_{,a}\dot Q^a=0,
\end{equation}
\begin{equation}
\label{bartheta-eqn}
\left(w(\mu,Q)\dot\theta\right)^.=0,
\end{equation}
\begin{equation}
\label{theta-eqn}
\left(w(\mu,Q)\dot{\bar\theta}\right)^.=0.
\end{equation}
In the operator approach the same equations are considered as equations
for operators with canonical commutation relations. (The introduction of the
latter ones is ensured by phase space extension). The construction of the
quantum theory of the Bianchi-IX model is described in Sec.\,\ref{Quant.dyn};
now we discuss the so-called ``conditionally-classical" model, where the
quantities appearing in (\ref{Seff}) are conditionally treated as classical
$c$-numerical and Grassmannian functions. The set (\ref{Q-eqn}) --
(\ref{theta-eqn}) differs from classical equations of general relativity
by the presence of ghosts and the inclusion of a gauge condition in the
number of the equations of motion (as a result, the Lagrange multiplier $\pi$
becomes a dynamical variable).

An important property of the set of equations is its {\em completeness}
enabling to formulate the Cauchy problem (see \cite{Git}) and to obtain
unambiguous extremals of the action. The equations are degenerate on
the subset of variables of the classical Einstein theory corresponding to a
trivial solution for ghosts and the parameter $\pi$, the ambiguity of their
solution does not enable one to define a path integral; solutions to such a
set of equations are called singular in a mathematical language. It is
obvious from (\ref{Q-eqn}) -- (\ref{theta-eqn}) that the explicit
substitution of trivial solutions for ghosts and the Lagrange multiplier
$\pi$ turns one back to the gauge-invariant classical set of equations.
But, being incomplete, the classical set of equations requires eliminating
coordinate effects, as it is done by special asymptotic boundary conditions
in the S-matrix approach that cannot be applied to a quantum version of the
discussed model. (It will be shown in Sec.\,\ref{WDWprobl} that wave
functions  calculated in the case of ``frozen out" ghost and gauge degrees
of freedom appear to be incorrect).

We construct a Hamiltonian dynamics in extended phase space by introducing
canonical momenta
$$
P_a=\frac{\partial L}{\partial\dot Q^a},\quad
P_0=\frac{\partial L}{\partial\dot\mu}=\pi,\quad
\bar\rho=\frac{\partial L}{\partial\dot\theta},\quad
\rho=\frac{\partial L}{\partial\dot{\bar\theta}}.
$$
The corresponding Hamiltonian is
\begin{equation}
\label{Hamilt}
H=P_a\dot Q^a+\pi\dot\mu+\bar\rho\dot\theta+\dot{\bar\theta}\rho-L
 =\frac12G^{\alpha\beta}P_{\alpha}P_{\beta}
 +\frac1{v(\mu,Q)}U(Q)
 -\frac i{w(\mu,Q)}\bar\rho\rho,
\end{equation}
where $\alpha=(0,a),\;Q^0=\mu$,
$$
G^{\alpha\beta}=\frac1{v(\mu,Q)}\left(
\begin{array}{cc}
f_{,a}f^{,a}&f^{,a}\\
f^{,a}&\gamma^{ab}
\end{array}
\right).
$$
The Hamiltonian (\ref{Hamilt}) gives the canonical equations
\begin{eqnarray}
\label{P-Ham}
\dot P_a&=&
  \frac1{v^2(\mu,Q)}v_{,a}\left[\frac12\left(P_bP^b
   +2\pi f_{,b}P^b+\pi^2f_{,b}f^{,b}\right)+U(Q)\right]\nonumber\\
&-&\frac1{v(\mu,Q)}\left[\pi f_{,ab}\left(P^b
  +\pi f^{,b}\right)+U_{,a}\right]
 -\frac i{w^2(\mu,Q)}w_{,a}\bar\rho\rho;\\
\label{Q-Ham}
\dot Q^a&=&
  \frac1{v(\mu,Q)}\left(P^a+\pi f^{,a}\right);\\
\label{pi-Ham}
\dot\pi&=&
  \frac1{v^2(\mu,Q)}v_{,\mu}\left[\frac12\left(P_aP^a
   +2\pi f_{,a}P^a+\pi^2f_{,a}f^{,a}\right)+U(Q)\right]
 -\frac i{w^2(\mu,Q)}w_{,\mu}\bar\rho\rho;\\
\label{mu-Ham}
\dot\mu&=&
  \frac1{v(\mu,Q)}f_{,a}\left(P^a+\pi f^{,a}\right);\\
\label{barrho-Ham}
\dot{\bar\rho}&=&0;\\
\label{theta-Ham}
\dot\theta&=&-\;\frac i{w(\mu,Q)}\rho;\\
\label{rho-Ham}
\dot\rho&=&0;\\
\label{bartheta-Ham}
\dot{\bar\theta}&=&-\;\frac i{w(\mu,Q)}\bar\rho.
\end{eqnarray}

It is easy to see that (\ref{P-Ham}) -- (\ref{bartheta-Ham})
are equivalent to the set of Eqs.\,(\ref{Q-eqn}) -- (\ref{theta-eqn});
the constraint (\ref{mu-eqn}) and the gauge condition (\ref{pi-eqn})
acquiring the status of Hamiltonian equations (\ref{pi-Ham}),
(\ref{mu-Ham}).

The primary (according to Dirac's terminology) constraint $\pi=0$ does not
appear in this approach because of the special choice of a gauge condition
(\ref{mudot}); the secondary constraint is modified as a consequence of
the extension of phase space. Once again, one can use trivial solution for
$\pi$ and ghosts to eliminate gauge-dependent terms. Another detail of the
Dirac approach is that gauge variables serve as Lagrange multipliers of
constraints. This makes one restrict the class of admissible parametrizations,
\begin{equation}
\label{mu-u}
v(\mu,Q)=\frac{u(Q)}{\mu}.
\end{equation}
Making use of the trivial solutions for $\pi$ and ghosts reduces
Eq.\,(\ref{pi-Ham}) in the class of parametrization (\ref{mu-u}) to
Dirac's secondary constraint,
\begin{equation}
\label{Dir-constr}
{\cal T}=\frac1{2u(Q)}P_aP^a+\frac1{u(Q)}U(Q)=0.
\end{equation}
The physical Hamiltonian in this class of parametrizations is proportional
to the secondary constraint, $H_0=\mu{\cal T}$. The constraint ${\cal T}$,
in its turn, does not depend on a gauge variable.

It is generally accepted that the trivial solutions for $\pi$ and ghosts
can be singled out by asymptotic boundary conditions (if one ignores
the problem of Gribov's copies). From this viewpoint asymptotic boundary
conditions lead to Dirac's constraints $\pi=0,\quad {\cal T}=0$.

The investigation of the set of equations (\ref{P-Ham}) --
(\ref{bartheta-Ham}) shows that there exist a conserved quantity
\begin{equation}
\label{BRSTgen}
\Omega
 =w(Q,\mu)\;\pi\dot\theta-H\theta
 =-\;i\;\pi\rho-H\theta.
\end{equation}
$\Omega$ generates the following transformations of variables in extended
phase space:
\begin{eqnarray}
\label{Q-tr}
\delta Q^a&=&
 \{Q^a,\Omega\}\bar\epsilon
 =-\;\frac{\partial H}{\partial P_a}\theta\bar\epsilon
 =-\;\dot Q^a\theta\bar\epsilon;\\
\label{mu-tr}
\delta\mu&=&
 \{\mu,\Omega\}\bar\epsilon
 =-\;i\rho\bar\epsilon-\;\frac{\partial H}{\partial\pi}\theta\bar\epsilon
 =w(\mu,Q)\dot\theta\bar\epsilon-\;\dot\mu\theta\bar\epsilon;\\
\label{theta-tr}
\delta\theta&=&
 \{\theta,\Omega\}\bar\epsilon=0;\\
\label{bartheta-tr}
\delta\bar\theta&=&
 \{\bar\theta,\Omega\}\bar\epsilon=-\;i\pi\bar\epsilon;\\
\label{P-tr}
\delta P_a&=&
 \{P_a,\Omega\}\bar\epsilon
 =\frac{\partial H}{\partial Q^a}\theta\bar\epsilon
 =-\;\dot P_a\theta\bar\epsilon;\\
\label{pi-tr}
\delta\pi&=&
 \{\pi,\Omega\}\bar\epsilon
 =\frac{\partial H}{\partial\mu}\theta\bar\epsilon
 =-\;\dot\pi\theta\bar\epsilon;\\
\label{barrho-tr}
\delta\bar\rho&=&
 \{\bar\rho,\Omega\}\bar\epsilon=H\bar\epsilon;\\
\label{rho-tr}
\delta\rho&=&
 \{\rho,\Omega\}\bar\epsilon=0.
\end{eqnarray}

The transformations (\ref{Q-tr}) -- (\ref{bartheta-tr}) are
BRST-transformations in the Lagrangian formalism.
This enables one to associate $\Omega$ with the BRST generator.

Dirac's constraints generate transformations for $Q^a,\;\mu$ that differ
from (\ref{Q-tr}), (\ref{mu-tr}). The difference in groups of
transformations results in the nonequivalence of the Lagrangian formulation
and the Hamiltonian one based on the set of constraints $\pi,\;{\cal T}$.

\sect{Comparison with the BFV approach}
\label{BFV}
It is interesting to compare these results with those obtained by a direct
application of the BFV approach.

The BFV method inherits some features of the Dirac approach, in particular,
one should restrict the class of admissible parametrizations (\ref{mu-u}).
We have the full set of constraints, ${\cal G}_{\alpha}=(\pi,\;{\cal T})$.
The BRST-generator is
\begin{equation}
\label{BFV-gen}
\Omega_{BFV}=\eta^{\alpha}{\cal G}_{\alpha}={\cal T}\theta-i\pi\rho.
\end{equation}
Taking a gauge-fixing function in the standard form,
$$
\bar\psi=i\bar\theta\chi(Q,P)+\bar\rho\mu,
$$
one obtains the effective action
\begin{eqnarray}
S_{BFV}&=&
 \!\int\!dt\,\biggl[
   P_a\dot Q^a+\pi\dot\mu+\dot\rho\bar\theta+\dot\theta\bar\rho
  +\{\bar\psi,\Omega\}\biggr]\nonumber\\
\label{S_BFV}
&=&\!\int\!dt\,\biggl[
   P_a\dot Q^a+\pi\dot\mu+\dot\rho\bar\theta+\dot\theta\bar\rho
  +i\bar\theta\{\chi,{\cal T}\}\theta
  -i\bar\rho\rho-\mu{\cal T}-\pi\chi(Q,P)\biggr].
\end{eqnarray}
It is easy to see that the effective action (\ref{S_BFV}) leads to a set of
equations which differs from (\ref{P-Ham}) -- (\ref{bartheta-Ham}).

The mathematical reason why the two formulations of Hamiltonian dynamics
are different is that the transformations generated by the constraint
(\ref{Dir-constr}) do not coincide with gauge transformations in Lagrangian
formalism. Correspondingly, the BRST charge (\ref{BFV-gen}) generates
transformations which differ from those generated by (\ref{BRSTgen}).
The form of transformations determines the structure of ghost Lagrangian,
so different groups of transformations lead to nonequivalent formulations
in extended phase space. The situation is typical for the theory of gravity.

The following may serve as a ``circumstantial evidence'' in favor of
the method of constructing Hamiltonian dynamics in EPS presented in
Sec.\,\ref{dynamics}. The path integral approach, in contrast of
canonical quantization, does not require to construct the Hamiltonian form
of the theory at the classical level. Hamiltonian operator is obtained
when deriving a Schr\"odinger equation from the path integral with
the action in Lagrangian form. It is remarkable that the application of
this procedure to the path integral with the effective action (\ref{Seff})
yields the Hamiltonian operator corresponding to the Hamiltonian
(\ref{Hamilt}) (see Sec.\,\ref{Quant.dyn}).

The two formulations of Hamiltonian dynamics in extended phase space could
enter into agreement in the class of parametrizations (\ref{mu-u}) in a
gauge-invariant sector which supposedly can be singled out by asymptotic
boundary conditions.

For a physical system possessing asymptotic states, neither of
the two formulations seems to have any advantages. However, being applied
to a system without asymptotic states, as a closed universe is, they may
give different results. In particular, the demand of BRST-invariance of
state vectors, $\Omega\,|\Psi\rangle=0$, where $\Omega$ is given by
(\ref{BRSTgen}), may not lead to the Wheeler -- DeWitt equation.
We shall return to this point in Sec.\,\ref{WDWprobl} after constructing
the quantum version of Hamiltonian dynamics in EPS.

\sect{The role of gravitational vacuum condensate}
\label{GVCrole}
An important feature of the conditionally-classical model is the
presence of a gravitational vacuum condensate (see Sec.\,8), its
quasi-energy-momentum tensor (quasi-EMT) being determined by a
parametrization and a gauge condition. To simplify further calculations
it is convenient to use an exponential parametrization
\begin{equation}
\label{exp.par}
v(\mu,Q)=\exp(\zeta(\mu,Q)).
\end{equation}

In a simple case, when the gauge function $f(Q^a)$ depends only on $Q^1=q$
and $v=\exp(\zeta(\mu,q))$, the quasi-EMT is isotropic:
$$
T_{\mu(obs)}^{\nu}=\mathop{\rm diag}\nolimits(\varepsilon_{(obs)},\,
 -p_{(obs)},\,-p_{(obs)},\,-p_{(obs)}),
$$
$$
\varepsilon_{(obs)}=-\frac{\dot\pi}{2\pi^2(\zeta_{,\mu})_k}
 \exp(\zeta_k-3q),
$$
$$
p_{(obs)}=\left\{1
 -\frac23\left[(\zeta_{,q})_k+(\zeta_{,\mu})_kf_{,q}\right]\right\}\varepsilon,
$$
where an index $k$ denotes that the substitution $\mu=f(Q^a)+k$ has been made;
Eq.\,(22), Sec.\,8, reads
$$
(\zeta_{,\mu})_k^{-1}\dot{\pi}=E_k.
$$
So, the GVC is a continual medium with the equation of state essentially
depending on a parametrization and a gauge, the latter two thus being
cosmological evolution factors. In other words, one of the main tasks of
theoretical cosmology appears to be finding out laws limiting a number of
admissible parametrizations and gauges, i.e. the investigation of the GVC
nature. Today we do not know how to approach this problem, but the
results of the present work, to all appearance, inevitably lead to it as a
constituent of the quantum measurement problem, mentioned in the report of
Penrose \cite{Pen} in the sense of finding new approaches to the construction
of quantum theory of gravity%
\footnote{The complexity of the problem, the
impossibility to solve it within the
framework of the conceptions of modern theoretical physics, in our opinion,
lies in the following. The introduction of a gauge condition even in the
classical theory of gravity is an operation establishing the integrity of the
system ``a physical object + observation means". The integrity consists in the
fact that space-time dynamics of gravitational and matter fields looks
differently in various reference systems. The possibility to extract a
gauge-invariant information in the classical theory is operationally ensured, 
according to its well-known conception, by that interactions between an
object and observation means can be made as weak as one would like. In
quantum theory (QT) the integrity of a system is established by introducing
commutation relations. The specificity of quantum theory of gravity is that we
try to realize the same physical idea, the idea of integrity, by means of
two {\em independent} mathematical operations which are  
parametrization-and-gauging and quantization. Apparently, in a
future theory it is necessary to have a physical principle and a formalism
that would unify these two operations.}.

Below we shall illustrate the role of the GVC in the cosmological evolution
in a simple case of a parametrization and a gauge allowing to
obtain an exact solution to the conditionally-classical set of 
Eqs.\,(\ref{Q-eqn}) -- (\ref{theta-eqn}), which an appropriate exact
quantum solution turns into in a semiclassical limit.

\sect{The conditionally-classical exact solution}
\label{Class.exact}
Taking the parametrization and the gauge 
\begin{equation}
\label{simplegauge}
v=\exp(\mu);\quad
\mu=k
\end{equation}
($k={\rm const}$), one can obtain an exact particular solution to
Eqs.\,(\ref{Q-eqn}) -- (\ref{theta-eqn}) with $Q^2=\varphi=0$. It is
the same gauge that was used and discussed in Sec.\,8. (see (12)). The
existence of the particular solution $\varphi=$0 gives the formal opportunity
to consider the model without this degree of freedom.

Under the condition (\ref{simplegauge}) the ghost variables vanish from
Eqs.\,(\ref{Q-eqn}) -- (\ref{pi-eqn}), and the
latter form a closed set concerning the physical variables; the
state equation of the GVC becomes extremely hard,
\begin{equation}
\label{GVC_state_eqn}
p=\varepsilon=-\frac{\dot{\pi}}{2\pi^2}\exp(k-3q),
\end{equation}
\begin{equation}
\label{lamb,E}
\dot{\pi}=E.
\end{equation}

Note that conditionality of the classical approach is shown here in
the presence of ghosts in the integral of motion (\ref{lamb,E}) 
$$
\dot{\pi}=\dot{\lambda}-\dot{\bar{\theta}}\dot{\theta},
$$
i.e. forms constructed on Grassmannian variables appear as parameters of the
theory.

Let us turn to the case of a single massless linear scalar field and put
$$
U_s(\phi)=0,
$$
in (\ref{U,Ug,Us}). Now we have the simple equation for $Q^4=\phi$
$$
\ddot{\phi}=0,
$$
thus,
\begin{equation}
\label{dotphi,Cs}
\dot{\phi}=C_s={\rm const}.
\end{equation}

The scalar field behaves as a medium with a positive energy density and with
an extremely hard equation of state
$$
p_{(scal)}=\varepsilon_{(scal)}\propto\exp(-3q)\dot{\phi}^2=C^2_s\exp(-3q)
$$
like that of the GVC (\ref{GVC_state_eqn}).

As one can see, in the present model the Universe is filled with the
two-component medium described by the parameters $E$
and $C_s$. Below we will show that the relation between the two parameters
essentially affects cosmological evolution at the quantum stage of the 
Universe existence as well as at the semiclassical one. Here is the
difference between our consideration and the usual investigation of the
Bianchi-IX model in general relativity.

The equations for $q,\,\chi$ take the form:
$$
\ddot{q}-\frac43\left[\exp(2q-4\chi)-4\exp(2q-\chi)\right]=0,
$$
$$
\ddot{\chi}-\frac43[2\exp(2q-4\chi)-2\exp(2q-\chi)]=0.
$$
Integration is simplified with the substitution
\begin{equation}
\label{z1,z2}
z_1=2q-4\chi,\quad
z_2=2q-\chi;
\end{equation}
after replacing
$$
t\to e^{-k}t
$$
the solution is written in the form
\begin{equation}
\begin{array}{l}
\label{solutions}
\exp\left(q-\displaystyle\frac12\chi\right)
 =\displaystyle\frac{\alpha}{\cosh[2\alpha(t-t_0)]},\\
\exp\left(q-2\chi\right)
 =\rule{0pt}{20pt}\displaystyle\frac{\beta}{\cosh[2\beta(t-t_1)]},
\end{array}
\end{equation}
where $\alpha,\,\beta,\,t_0,\,t_1$ 
are the integration constants. Without loss of generality by shifting
the origin of time coordinate one can put $t_1=0$.
For the metric (\ref{ds}) -- (\ref{eta_ab}) one finds:
$$
a^2=b^2=\frac14\exp(q+\chi)
 =\frac{a^2\cosh(2\beta t)}{4\beta\cosh^2[2\alpha(t-t_0)]};
$$
$$
c^2=\frac14\exp(q-2\chi)=\frac{\beta}{4\cosh(2\beta t)}.
$$
From the constraint equation (\ref{mu-eqn}) with $\phi=0$ and
(\ref{dotphi,Cs}) it follows:
$$
\frac12\left[\dot{\chi}^2-\dot{q}^2\right]+\frac23\left[\exp(2q-4\chi)
 -4\exp(2q-\chi)\right]=E_k-\frac12C^2_s,
$$
where $E_k={\rm e}^kE$; hence, in turn,
\begin{equation}
\label{alpha,beta-invar}
\alpha^2-\frac14\beta^2=-\frac38\left[E_k-\frac12C^2_s\right].
\end{equation}
The dynamics of the model depends qualitatively on a relation between
$C_s$ and $E_k$ .

1. Empty space (there is neither scalar field, $C_s=0$, nor GVC, $E_k=0$);
$\alpha=\displaystyle\frac{\beta}2$.

In the limit $t=\pm\infty$
$$
a^2=b^2=\frac{\beta}8;\quad
c^2=0,
$$
i.e. the metric (\ref{ds}) asymptotically takes the form
$$
ds^2=(\beta c)^2dt^2-\frac{\beta}8(d\vartheta^2+\sin^2\vartheta d\varphi^2),
$$
$$
\vartheta=x^1,\quad
\varphi=x^2.
$$

When reaching singularity in one of the dimensions, two others form a
stationary space of constant curvature. Here one deals with a regime of
dynamical compactification, a space with simple topology being compactified.

2. Space is filled with the medium having a positive energy density:
$E_k<\displaystyle\frac12C_s^2;\quad \alpha>\displaystyle\frac12\beta$.

For $\alpha=\beta,\;t_0=0$ the model is isotropic.

For $\alpha=\beta,\;t_0\ne0$ the model is anisotropic, but the singularity
has an isotropic nature.

For $\displaystyle\frac12\beta<\alpha<\beta$ in a pre-singular state 
$a^2=b^2\gg c^2$, i.e. (2 + 1)-dimensional space-time arises where
2-space has a constant curvature.

For $\alpha>\beta$ in a pre-singular state $c^2\gg a^2=b^2$, however, the
model is not reduced to a space of less dimensions.

In all the cases for $E_k<\displaystyle\frac12C_s^2$ space at singularity is 
contracted to a point.

3. Space is filled with the medium having a negative energy density: 
$E_k>\displaystyle\frac12C_s^2;\quad \alpha<\displaystyle\frac12\beta$.

At $t=\pm\infty$ the third space dimension is compactified ($c^2\to0$), and
the remaining two-dimensional space of constant curvature infinitely
expands. In the special case $\alpha=\displaystyle\frac{\beta}{4}$
the scale factor $a=b$ increases exponentially in proper time.

So, the GVC affecting coupling between the constants $\alpha$ and $\beta$
through the controlling parameter $E_k$ determines a cosmological scenario
which may contain the following phenomena:
\newcounter{phen}
\begin{list}{\arabic{phen})}%
 {\topsep=3pt
  \itemsep=3pt\parsep=0pt
  \usecounter{phen}}
\item cosmological expansion and contraction of space;
\item cosmological singularity;
\item compactification of space dimensions;
\item asymptotically stationary space of less dimensions; 
\item inflation of the Universe.
\end{list}
One can see that even such a simplified model reveals a number of effects
which seem to be probable from the standpoint of modern cosmological ideas.
The introduction of the GVC to the theory enlarges the number of possible
cosmological scenarios, a concrete value of the parameter $E_k$ being formed
at the quantum stage of the Universe existence.

\sect{Quantum dynamics} 
\label{Quant.dyn}
It is essentially important to note that in the EPS formalism a dynamical
Schr\"odinger equation is a direct and unambiguous consequence of
canonical quantization procedure by no means depending on our concepts about
gauge invariance or noninvariance of the theory. Really, a Schr\"odinger
equation can be derived from the quantum canonical equations
\begin{equation}
\label{H,X}
\dot{X}=i\,[H,X],
\end{equation}
written in the matrix form,
$$
\langle\Psi_1|\dot{X}|\Psi_2\rangle
 \equiv\frac{\partial}{\partial t}\langle\Psi_1|X|\Psi_2\rangle
 =\left\langle\frac{\partial\Psi_1}{\partial t}\right|X|\Psi_2\rangle
 +\langle\Psi_1|X\left|\frac{\partial\Psi_2}{\partial t}\right\rangle,
$$
$$
\langle\Psi_1|\,[H,X]\,|\Psi_2\rangle
 =\left(\langle\Psi_1|H\right)\left(X|\Psi_2\rangle\right)
 -\left(\langle\Psi_1|X\right)\left(H|\Psi_2\rangle\right),
$$
where $|\Psi_1\rangle,\;|\Psi_2\rangle$ are arbitrary state vectors.
Joining the latter two formulae into the equation corresponding to
(\ref{H,X}) and taking into account the arbitrariness of the state vectors
$|\Psi_1\rangle,\;|\Psi_2\rangle$ and
$X|\Psi_1\rangle,\;X|\Psi_2\rangle$, one comes to mutually conjugate
dynamical Schr\"odinger equations with the Hamiltonian operator (\ref{Hamilt})
defined in the EPS:
$$
i\,\frac{\partial|\Psi\rangle}{\partial t}=H|\Psi\rangle,\quad
-i\,\frac{\partial\langle\Psi|}{\partial t}=\langle\Psi|H.
$$

A dynamical Schr\"odinger equation, surely, can also be obtained in the
path integral formalism having certain advantages over the operator one.
In the latter, as is generally known, the operator ordering problem is not
resolvable. When deriving a Schr\"odinger equation from a path integral,
ordering turns to be bound up with a way of a final definition of the path
integral as the limit of a multiple integral and with a choice of a gauge
variable parametrization. The parametrization choice determines a path
integral measure as well, the latter being identical with the measure of a
normalizing integral -- probability measure. Let us consider a path integral
for a wave function in the coordinate representation. Such a wave function,
according to the stated above, is defined on the extended configurational
space with the coordinates $Q^a,\,\mu,\,\theta,\,\bar{\theta}$:
$$
\mbox{\hbox to \textwidth{$
\Psi(Q^a,\,\mu,\,\theta,\,\bar{\theta};\,t)
 =\displaystyle\int\langle\,Q^a,\,\mu,\,\theta,\,\bar{\theta};\,t\,|\,
    Q^a_{(0)},\,\mu_{(0)},\,\theta_{(0)},\,\bar{\theta}_{(0)};\,t_0\,\rangle
   \Psi(Q^a_{(0)},\,\mu_{(0)},\,\theta_{(0)},\,\bar{\theta}_{(0)};\,t_0)
   \times
$ \hfil}}
$$
\begin{equation}
\label{WF-PI}
   \times M\left(Q^a_{(0)},\,\mu_{(0)}\right)
    d\theta_{(0)}\,d\bar{\theta}_{(0)}\,d\mu_{(0)}\prod_b dQ^b_{(0)}.
\end{equation}
The transition amplitude, appearing in (\ref{WF-PI}),
{\samepage
$$
\mbox{\hbox to\textwidth{$
\langle\,Q^a,\,\mu,\,\theta,\,\bar{\theta};\,t\,|\,
 Q^a_{(0)},\,\mu_{(0)},\,\theta_{(0)},\,\bar{\theta}_{(0)};\,t_0\,\rangle=
$ \hfil}}
$$
$$
 =C\int\exp\left[iS(t,\,t_0)\right]
  \prod_{t_0<\tau<t}M\left(Q^a_{(\tau)},\,\mu_{(\tau)}\right)
  d\mu_{(\tau)}\,d\theta_{(\tau)}\,d\bar{\theta}_{(\tau)}
  \prod_bdQ^b_{(\tau)}\,d\pi_{(\tau)}\,d\pi_{(t)},
$$}
is given by the gauged action (\ref{Seff}).

As is well known, a path integral is not defined in internal terms.
Proceeding from the standard treatment we shall consider it as the limit
at $\epsilon_i\to0$ of the following integral:
$$
\mbox{\hbox to\textwidth{$
\Psi^{(N)}(Q^a,\,\mu,\,\theta,\,\bar{\theta})
 =C\displaystyle\int\exp\left\{i\sum^N_{i=1}S(t_i,t_{i-1})\right\}
  \Psi^{(0)}(Q^a,\,\mu,\,\theta,\,\bar{\theta})\times
$ \hfil}}
$$
$$
  \times\prod^{N-1}_{i=0}M\left(Q^a_{(i)},\,\mu_{(i)}\right)
  d\mu_{(i)}\,d\theta_{(i)}\,d\bar{\theta}_{(i)}
  \prod_bdQ^b_{(i)}\,d\pi_{(i+1)},
$$
where $t_i-t_{i-1}=\epsilon_i$,
$$
S(t_i,\,t_{i-1})\approx\epsilon_i\biggl\{\displaystyle\frac12
 \exp(\zeta_{(i)})\,\dot{Q}^a_{(i)}\,\dot{Q}_a^{(i)}
 -\exp(-\zeta_{(i)})\,U(Q^a_{(i)})
 +\pi_{(i)}\left[\dot{\mu}_{(i)}-\dot{f}(Q^a_{(i)})\right]
 +\frac i{\zeta_{,\mu(i)}}\dot{\bar{\theta}}_{(i)}\dot{\theta}_{(i)}\biggr\}.
$$

The further process of the path integral definition consists in choosing an
approximation for the paths between $t_{i-1}$ and $t_i.$ The standard
procedure, which we still do not see any reason to deviate from, prescribes
approximating by classical paths. In our case it means applying the
equations of motion (\ref{Q-eqn}) -- (\ref{theta-eqn}). To calculate the
path integral it is necessary to express all the velocities in terms of the
values of coordinates at the end points of a time step $\epsilon_i$ by means
of the equations of motion \cite{FH,Cheng}, i.e. to solve the Cauchy
problem that could be put only on the basis of the complete set of
equations (\ref{Q-eqn}) -- (\ref{theta-eqn})%
\footnote{It should be emphasized that in the case of
the lack of asymptotic states there exist no other way compatible with the
principles of QT. The known attempts (see, for example,
\cite{BP}) to derive gauge-invariant equations make use of the boundary
conditions of the type
$$
\theta(t_0)=\theta(t)=0,
$$
$$
\bar{\theta}(t_0)=\bar{\theta}(t)=0,
$$
$$
\pi(t_0)=\pi(t)=0.
$$
The application of these conditions in the path integral at $t-t_0=\epsilon\to0$ when
deriving the Schr\"odinger equation and the Wheeler -- DeWitt equation eliminates from the path integral all the structures
associated with a gauge and restore the initial divergent integral with the
gauge-invariant action. Appealing to these conditions does not make sense
unless one points out the way to eliminate the contribution of the coordinate
effects in the path integral owing to the degeneracy of the Cauchy problem for the
action extremals. Such boundary conditions are not applied practically even as
asymptotic conditions in S-matrix problems where the {\em vacuum} of ghosts
and of 3-vector and 3-tensor gravitons is given but {\em not fixed values} of
ghost fields and Lagrange multipliers.
\label{zero-boundary}}.
When solving the problem with the required accuracy the expansions in series of
$\epsilon_i$ powers are used,
$$
\dot{Q}^a_{(i)}=\frac1{\epsilon_i}\left(Q^a_{(i)}-Q^a_{(i-1)}\right)
 +\displaystyle\frac12\epsilon_i\ddot{Q}^a_{(i)}
 -\frac16\epsilon^2_i\stackrel{...}{Q}^a_{(i)}+\cdots
$$
and the analogues ones for $\mu$ and ghosts. Then the higher derivatives with
respect to time are consequently expressed in terms of the first derivatives
through the equations of motion. As a result, all the velocities appear in the
form of power series of the differences
$$
q^a_{(i)}=Q^a_{(i)}-Q^a_{(i-1)},
$$
$$
m_{(i)}=\mu_{(i)}-\mu_{(i-1)},
$$
$$
\eta_{(i)}=\theta_{(i)}-\theta_{(i-1)},
$$
$$
\bar{\eta}_{(i)}=\bar{\theta}_{(i)}-\bar{\theta}_{(i-1)}.
$$
In particular, the gauge condition is approximated by the following:
$$
\dot{\mu}_{(i)}-\dot{f}_{(i)}=\frac1{\epsilon_i}\left(\mu_{(i)}-\mu_{(i-1)}
 -f_{(i)}+f_{(i-1)}\right)
 =\frac1{\epsilon_i}\left(m_{(i)}-f_{,a}^{(i)}q^a_{(i)}
 +\displaystyle\frac12f_{,a,b}^{(i)}\,q^a_{(i)}q^b_{(i)}+\cdots\right).
$$

When deriving the Schr\"odinger equation it is sufficient to consider a one-step time
interval $t-t_0=\epsilon$. The further procedure consist of substituting
the expansions for the velocities into the exponent $\exp(iS)$, developing
it as series in  $\epsilon$, developing the measure
$M\left(Q^a(t_0),\,\mu(t_0)\right)=M\left(Q^a(t)-q^a,\,\mu(t)-m\right)$
as series in $q^a,\,m$, and developing the wave function $\Psi^{(0)}$ as series in
$q^a,\,m,\,\eta,\,\bar{\eta}$. After performing simple integrations over
$\pi$ and $m$, the path integral is reduced to Gaussian quadratures over the ghost
and physical variables. Then by the standard Feynman method in first order
one obtains the Schr\"odinger equation
\begin{equation}
\label{SE1}
i\,\frac{\partial\Psi(Q^a,\mu,\theta,\bar{\theta};\,t)}{\partial t}
 =H\Psi(Q^a,\,\mu,\,\theta,\,\bar{\theta};\,t)
\end{equation}
the zero-order terms will give the constraints
between the measure $M$, step $\epsilon$ and parametrization function $\zeta$:
\begin{equation}
\label{eps,M-coupl}
\frac1{\epsilon \zeta_{,\mu}}
 \left(\epsilon\,{\rm e}^{-\zeta}\right)^{\textstyle\frac K2}M={\rm const}.
\end{equation}
The requirement for the Hamiltonian to be Hermitian gives raise to another
constraint between the measure and the parametrization,
\begin{equation}
\label{zeta,M-coupl}
M={\rm const}\cdot \zeta_{,\mu}\exp\left(\frac{K+3}2\,\zeta\right).
\end{equation}
The independence of the parameter $\epsilon$ on the variables $Q^a,\,\mu$
follows from (\ref{eps,M-coupl}), (\ref{zeta,M-coupl}).

The Hamiltonian in the Schr\"odinger equation obtained by the path integral method can be presented in
the form
\begin{equation}
\label{H_full}
H=-i\,\zeta_{,\mu}\frac{\partial}{\partial\theta}
  \frac{\partial}{\partial\bar{\theta}}
 -\frac1{2M}\frac{\partial}{\partial Q^{\alpha}}\tilde G^{\alpha\beta}
  \frac{\partial}{\partial Q^{\beta}}+{\rm e}^{-\zeta}(U-V),
\end{equation}
where $M$ is defined by the formula (\ref{zeta,M-coupl}),
$\tilde{G}^{\alpha\beta}=M\,G^{\alpha\beta}$,
\begin{equation}
\label{V}
V=-\frac3{12}\frac{(\zeta_{,\mu})^a(\zeta_{,\mu})_a}{\zeta_{,\mu}^2}
 +\frac{(\zeta_{,\mu})^a_a}{3\zeta_{,\mu}}
 +\frac{K+1}{6\zeta_{,\mu}}\,\zeta_a(\zeta_{,\mu})^a
 +\frac1{24}\left(K^2+3K+14\right)\,\zeta_a\zeta^a+\frac{K+2}6\,\zeta_a^a,
\end{equation}
$$
\zeta_a=\frac{\partial \zeta}{\partial Q^a}+f_{,a}\frac{\partial \zeta}{\partial\mu}.
$$

Let us compare the operator formalism and the path integral one. Taking into account
the form of a self-conjugate momentum operator in the presence of a
nontrivial measure,
\begin{equation}
\label{P-of-M}
P_{\alpha}=-i\left(\frac{\partial}{\partial Q_{\alpha}}
 +\frac{M_{,\alpha}}{2M}\right),
\end{equation}
it is easy to see that in the operator formalism Eq.\,(\ref{H_full})
corresponds to the Hamiltonian (\ref{Hamilt}) ordered according to
\begin{equation}
\label{H,order}
H=\frac12\sum_{i=1}^n a_i\xi_iP_{\alpha}\xi_i^{-2}G^{\alpha\beta}
 P_{\beta}\xi_i+{\rm e}^{-\zeta}U-i\zeta_{,\mu}\bar{\rho}\rho,
\end{equation}
$$
\sum_{i=1}^n a_i=1,
$$
with $\xi_i=(\zeta_{,\mu})^{r_i}\exp(s_i\zeta),\,n=2$; the comparison of the
Hamiltonians (\ref{H,order}) and (\ref{H_full}) gives 5 equations for
the 5 independent parameters $r_i,s_i,a_1$.

But, on account of the mentioned above definition ambiguity of the path integral, such an
ordering should not be treated as the only possible one. The ordering problem
as well as the problem of the choice of a gauge variable has no solution in
the QGD framework. The existence of the two problems indicates the
incompleteness of the theory. They have the common origin -- the lack of
understanding what is a measurement process in quantum gravity. Indeed,
the choice of parametrization and gauge in total determine an instrument
tuning (a time counting scale). The ordering problem arises from the operator
noncommutativeness which, in turn, is the consequence of unremovable
instrument affection on a physical system.

When dealing with the quantum physics domains available to experimental
investigation, the experiment itself indicates solutions to problems. But in
quantum gravity it is necessary to achieve a new level of understanding for
that, in other words, the question is about the creation of a quantum measurement
theory (see \cite{Pen,Penr}). In our opinion, the proposed modification
of QGD giving possibility to describe jointly a physical object and
OM may be considered as a first step on that way.

\sect{The structure of the general solution}
\label{GS}
The dynamical Schr\"odinger equation (\ref{SE1}) with the Hamiltonian
(\ref{H_full}) has the general solution which should be considered as the
quantum analogue of the solution to the Cauchy problem for
conditionally-classical equations (\ref{Q-eqn}) -- (\ref{theta-eqn}).
Now we will turn to the analysis of the general solution
structure that, as it is obvious in advance, is not gauge-invariant.

The general solution to the Schr\"odinger equation in the coordinate representation is a wave function
$\Psi=\Psi(Q^a,\,Q^0,\,\theta,\,\bar{\theta}\,t)$,
depending on time $t$, physical variables $Q^a$, the gauge variable
$Q^0\equiv\mu$, and ghost variables $\theta,\bar{\theta}$. Let us show that
making use of the Hamiltonian structure only one can establish the dependence
of the wave function on the variables $Q^0,\,\theta,\,\bar{\theta}$.

To begin with, note that in the class of gauges (\ref{mudot}) not depending
on time explicitly the general solution to the Schr\"odinger equation (\ref{SE1}) is expandable in
stationary state eigenfunctions satisfying the stationary Schr\"odinger equation
\begin{equation}
\label{station_SE}
H\Psi_n(Q^a,\,Q^0,\,\theta,\,\bar{\theta})
 =E_n\Psi_n(Q^a,\,Q^0,\,\theta,\,\bar{\theta}).
\end{equation}
One of the canonical equations, the gauge equation
$$
\left[H,\,Q^0-f(Q^a)\right]=0
$$
means the commutativeness of the Hamiltonian with the operator
$Q^0-f(Q^a)$ and, consequently, an arbitrary solution to
Eq.\,(\ref{station_SE}) can be presented in the form of a superposition of
this operator eigenstates $|k\rangle$,
$$
\{Q^0-f(Q^a)\}|k\rangle=k\,|k\rangle.
$$
The same concerns the general solution to (\ref{SE1}) as a superposition of stationary
states. In the coordinate representation
\begin{equation}
\label{k-vector}
|k\rangle=\delta\left(Q^0-f(Q^a)-k\right),
\end{equation}
so the general solution to the Schr\"odinger equation has the structure
\begin{equation}
\label{delta-factor}
\Psi(Q^a,\,Q^0,\,\theta,\,\bar{\theta};\,t)=\int
 \Phi_k(Q^a,\,\theta,\bar{\theta};\,t)\,\delta\left(Q^0-f(Q^a)-k\right)\,dk.
\end{equation}
Since there is no other (independent) integral of motion for the $Q^0$
variable, the functions (\ref{k-vector}) make the only basis depending on the 
gauge variable $Q^0$, i.e. the general solution to the dynamical Schr\"odinger equation inevitably has the
structure (\ref{delta-factor}). One may come to the same conclusion by
investigating the structure of the wave function in the path integral formalism.

The Schr\"odinger equation for $\Phi_k(Q^a,\,\theta,\,\bar{\theta};\,t)$ reads
\begin{equation}
\label{Phi_k-SE}
i\,\frac{\partial \Phi_k(Q^a,\theta,\bar{\theta};t)}{\partial t}
 =H_k\,\Phi_k(Q^a,\,\theta,\,\bar{\theta};\,t),
\end{equation}
$$
H_k=-i\,(\zeta_{,\mu})_k\frac{\partial}{\partial\theta}
 \frac{\partial}{\partial\bar{\theta}}
 -\displaystyle\frac12\exp(-\zeta_k)\left(
  \frac{\partial^2}{\partial Q^a\partial Q_a}
 +Z^a_k\frac{\partial}{\partial Q^a}\right)+\exp(-\zeta_k)(U-V),
$$
$$
Z^a_k=\frac{(\zeta_{,\mu})^a_k}{(\zeta_{,\mu})_k}+\frac{K+1}2\,\zeta^a_k,
$$
$V$ being defined by Eq.\,(\ref{V}),
$(\zeta^a)_k=\zeta_k^{,a}\equiv\partial \zeta_k/\partial Q_a$.

So, the wave function dependence on $\mu$ is determined by Eq.\,(\ref{delta-factor}). As
will be shown below, such a structure of the wave function under a certain restriction on
the $\Phi_k$ dependence on $k$ makes the normalizing integral over the variable
$\mu$ transformed into an integral over k to be convergent. As for the
dependence on the ghosts, it is strictly enough fixed by the Schr\"odinger equation in combination
with the usual demand of norm positivity. Indeed, in the general case the
wave function can be presented by the series in Grassmannian variables,
$$
\Phi_k(Q^a,\,\theta,\,\bar{\theta};\,t)
 =\Psi^0_k(Q^a,t)+\Psi^1_k(Q^a,t)\,\theta
 +\Psi^{\bar{1}}_k(Q^a,t)\,\bar{\theta}+\Psi^2_k(Q^a,t)\,\bar{\theta}\theta.
$$
After substituting into (\ref{Phi_k-SE}) one obtains the equations for the
components
\begin{equation}
\label{ghost-expan-SE1}
i\,\frac{\partial\Psi^0_k}{\partial t}=H^0_k\Psi^0_k-i\,(\zeta_{,\mu})_k\Psi^2_k,
\end{equation}
\begin{equation}
\label{ghost-expan-SE2}
i\,\frac{\partial\Psi^i_k}{\partial t}=H^0_k\Psi^i_k,\quad
i=1,\,\bar{1},\,2,
\end{equation}
$$
H^0_k=H_k+i\,(\zeta_{,\mu})_k\,\frac{\partial}{\partial\theta}
 \frac{\partial}{\partial\bar{\theta}},
$$
and the normalization condition imposes the constraints on these components: 
from the norm positivity demand it follows
$$
-i\int\left(\Psi^{0*}_k\Psi^2_k-\Psi^{2*}_k\Psi^0_k
 +\Psi^{\bar{1}*}_k\Psi^1_k-\Psi^{1*}_k\Psi^{\bar{1}}_k\right)
 \bar{\theta}\theta\,d\theta\,d\bar{\theta}>0.
$$
One of the consequences of the nonequality is $\Psi_k^2=0$, or $\Psi_k^0=0$,
or $\Psi_k^2=i\Psi_k^0$, and Eqs.\,(\ref{ghost-expan-SE1}),
(\ref{ghost-expan-SE2}) reduce all the three versions to the one,
$$
\Psi^0_k=\Psi^2_k=0,
$$
$$
\Psi^1_k=i\Psi^{\bar{1}}_k,
$$
so, finally,
\begin{equation}
\label{Phi-of-ghost}
\Phi_k(Q^a,\,\theta,\,\bar{\theta};\,t)=\Psi_k(Q^a,t)(\bar{\theta}+i\theta),
\end{equation}
where $\Psi_k(Q^a,t)$ is a solution to Eq.\,(\ref{ghost-expan-SE2}):
$$
i\,\frac{\partial\Psi_k(Q^a,t)}{\partial t}
 =-\frac1{2M_k}\frac{\partial}{\partial Q^a}\,M_k\exp(-\zeta_k\,)\gamma^{ab}\,
  \frac{\partial\Psi_k(Q^a,t)}{\partial Q^b}
 +\exp(-\zeta_k)\,(U-V_k)\,\Psi_k(Q^a,t),
$$
$$
M_k=(\zeta_{,\mu})_k\exp\left(\frac{K+3}2\,\zeta_k\right).
$$
The unitarity property of the wave function of a physical state takes the form
$$
\mbox{\hbox to\textwidth{$
\displaystyle\int\Psi^*_{k'}(Q^a,t)\,\delta(\mu-f(Q^a)-k')\,\Psi_k(Q^a,t)\,
 \delta(\mu-f(Q^a)-k)\,dk\,dk'\,M(Q^a,\mu)\,d\mu\prod_{a=1}^{K+3}dQ^a=
$ \hfil}}
$$
\begin{equation}
\label{norm.int}
 =\int\Psi^*_k(Q^a,t)\Psi_k(Q^a,t)\,dk\prod_adQ^a.
\end{equation}

Thus, the general solution (\ref{delta-factor}), (\ref{Phi-of-ghost}) to Eq.\,(\ref{SE1})
under the condition the $\Psi_k(Q^a,t)$ to be a sufficiently narrow packet
over $k$ is normalizable with respect to the gauge variable as well as to
the ghosts and the physical variables.

The peculiarity of the amplitude (\ref{Phi-of-ghost}) lies in the fact that
the theory does not control its dependence on the free parameter $k$. From
the standpoint of classical dynamic equations the parameter $k$ sets an
initial condition for the variable $\mu$ and by the same determines an
initial clock setting. In QT, however, there exist no
physical state with a fixed $k$ value. Really, the unitarity requirement (see
(\ref{norm.int})) allows the existence of a physical state represented by a
packet over $k$ narrow enough to fit a certain classical $\bar{k}$ value, but
not by a $\delta$-shaped packet. So, in the theory an additional degree of
freedom appears that could be named {\em an observer's degree of freedom}.
Unlike the quantum uncertainties associated with operator noncommutativeness,
QGD do not control even a width of a $k$-packet.

The general solution structure
\begin{equation}
\label{time-depend.WF}
\Psi(Q^a,\,Q^0,\,\theta,\,\bar{\theta};\,t)
 =\int\Psi(E_k;\,Q^a)\exp(-iE_kt)(\bar{\theta}+i\theta)\,
  \delta(\mu-f(Q^a)-k)\,dE_k\,dk,
\end{equation}
where $\Psi(E_k;\,Q^a)$ is a solution to the stationary equation
\begin{equation}
\label{station.phys.SE}
H^0_k\,\Psi_k(Q^a)=E_k\,\Psi_k(Q^a),
\end{equation}
{\em proves mathematically} all the statements of Sec.\,5: the wave function carries the
information on 1) a physical object, 2) OM, 3) correlations between a physical
object and OM. OM are represented by the factored part of the wave function -- by the
$\delta$-function of a gauge and by the ghosts; a physical object is described 
by the function $\Psi_k$; the correlations are present in the effective
potential $V_k$ which the solution depends on and also in the wave function time
dependence; after going over to the stationary states, they are present
in the effective potential and the spectrum $E_k$.

It is of principal significance to emphasize that the procedure of
constructing the wave function (\ref{time-depend.WF}) is the only strict mathematical
method to do it, by no way corresponding to the Wheeler -- DeWitt QGD. The question arises:
do Eq.\,(\ref{SE1}) and its solutions have any relation to the Wheeler -- DeWitt theory?
In other words, whether it is possible to extract such a physical part of the
(\ref{time-depend.WF}) that would satisfy the Wheeler -- DeWitt equation and could
be reasonably interpreted?

\sect{Gauge-invariant QGD\\
and the requirement of BRST-invariance of physical states}
\label{WDWprobl}
To begin with, let us discuss a possibility to construct a wave function,
corresponding to the Wheeler -- DeWitt QGD, by going over from the general solution to some particular
solution. As it is known, the transition to the Wheeler -- DeWitt QGD means the separation
of physical variable subspace from EPS. For this purpose it is not enough to
separate the physical part $\Psi_k(Q^a)$ from the general solution: one also has to banish
correlations between the properties of the physical object and those of OM.
The latter are given, first of all, by the GVC parameter. Hence, to eliminate
correlations, firstly, we put $E_k=0$. Secondly, making use of the noted in
Sec.\,\ref{Model} possibility to go over to any given gauge function
$f(Q^a)$ by means of transformation of a parametrization function $v(\mu,Q^a)$
it is necessary to pass to the gauge
\begin{equation}
\label{triv.gauge}
\mu=k.
\end{equation}
And, thirdly, the measure should be factored: $M=M_1(\mu)\,M_2(Q^a)$. In
view of Eq.\,(\ref{zeta,M-coupl}) the measure factorization requires, in turn,
factoring the parametrization function that means
\begin{equation}
\label{zeta_add}
\zeta(\mu,Q^a)=\zeta_1(\mu)+\zeta_2(Q^a),
\end{equation}
the conservation of additivity of the function (\ref{zeta_add}) when going over
to the gauge (\ref{triv.gauge}) imposing one more restriction on the
parametrization function: $\zeta_1$ should be linear in $\mu$,
\begin{equation}
\label{zeta1_lin}
\zeta_1(\mu)=A+B\mu.
\end{equation}
Under these conditions one obtains the Wheeler -- DeWitt equation for the physical part of
the wave function
\begin{equation}
\label{WDWeqn}
H_{ph}\Psi(Q^a)=0,
\end{equation}
\begin{equation}
\label{H_ph}
H_{ph}=-\frac1{2M_2}\frac{\partial}{\partial Q^a}\,M_2\exp(-\zeta_2)\,\gamma^{ab}
  \frac{\partial}{\partial Q^b}+\exp(-\zeta_2)\,U+\frac16{\bf R},
\end{equation}
\begin{equation}
\label{R}
{\bf R}=-\exp(-\zeta_2)\bigg[\frac14\left(K^2+3K+14\right)\zeta_2^{,a}
  \zeta^{\phantom{,a}}_{2,a}+\left(K+2\right)\zeta^{,a}_{2,a}\bigg].
\end{equation}
${\bf R}$ being the scalar curvature constructed on the metric
$G^{ab}=\exp(-\zeta_2)\gamma_{ab}$.

Eq.~(\ref{WDWeqn}) possesses all formal properties of the Wheeler -- DeWitt
equation including parametrization noninvariance and the lack of any visible
vestige of a gauge. {\it However, the described above method of deriving
Eq. (\ref{WDWeqn}) makes it apparent that in the gauge class (\ref{mu,f,k}),
any change of the parametrization function $\zeta(\mu,Q^a)$ is
mathematically equivalent to a new gauge. Hence, the generally known
parametrization noninvariance of the Wheeler -- DeWitt theory, as a matter
of fact, is the ill-hidden gauge noninvariance}. This circumstance has to
be taken into account when estimating the status of the Wheeler -- DeWitt
theory.

On the other hand, by origin, the only Hamiltonian eigenvalue $E_k=0$ fixed
by Eq.\,(\ref{WDWeqn}) is by construction a {\em line in a continuous
spectrum} of the Hamiltonian (\ref{H_ph}), hence the solution is
unnormalizable. In other words, on the way of formal singling out the
particular solution to Eq.\,(\ref{SE1}) one would fail to obtain a wave
function having a physical meaning generally adopted in QT%
\footnote{A positive solution to this problem needs, apart from a
discrete $H_{ph}$ spectrum, the presence of the line $E=0$ in it.}.

Another approach to the problem of the existence of gauge invariant states
is based on singling out BRST-invariant solutions. In this case the
procedure of derivation of the Wheeler -- DeWitt equation does not consist
in fixing an eigenvalue in a Hamiltonian spectrum, but in {\it reducing}
the spectrum {\it itself} to a single value by means of the BRST-invariance
requirement.

It is worth noticing that this requirement should be treated as an independent
postulate: the BRST invariance of the action cannot be a foundation for
introducing them. It is known that the invariance of an action under some
global transformations does not mean the invariance of state vectors; the
latter ones have to be just {\em covariant}, subjected to the appropriate
unitary transformations%
\footnote{For instance, in the standard QT of fields the Lorentz invariance
of an action leads to the Lorentz covariance of state vectors. The only
Loretz-invariant vector is a vacuum vector which is a particular solution
obtained from the general one.}.

In the BFV scheme a wave function of a physical state should
obey the superselection rules
\begin{equation}
\label{BRST1-constr}
\Omega_{BFV}\,|\Psi\rangle=0,
\end{equation}
where $\Omega_{BFV}$ is given by (\ref{BFV-gen}). The Wheeler -- DeWitt
equation
\begin{equation}
\label{WDW-constr}
{\cal T}\,|\Psi\rangle=0
\end{equation}
follows immediately from (\ref{BRST1-constr}) as a consequence of
the arbitrariness of BFV ghosts $\{\eta^{\alpha}\}$. This result is quite
natural, because the theory is constructed in such a way for operator
Dirac's constraints to be inevitably satisfied.

In the approach presented in this paper the sense of the requirement
\begin{equation}
\label{BRST2-constr}
\Omega\,|\Psi\rangle=0,
\end{equation}
where $\Omega$ is given by (\ref{BRSTgen}), is more questionable.
The generator $\Omega$ in this case cannot be presented as a combination
of constraints with infinitesimal parameters replaced by ghosts.

One may suppose that the condition (\ref{BRST2-constr}) together with
the quantum version of the primary constraint,
\begin{equation}
\label{pi-constr}
\pi\,|\Psi\rangle=0,
\end{equation}
lead to the Wheeler -- DeWitt equation. (At least, for a system with
asymptotic states one may expect that the primary constraint $\pi=0$ is
valid.) However, for any operator ordering, the additional condition
(\ref{pi-constr}) does not reduce (\ref{BRST2-constr}) to the Wheeler --
DeWitt equation (\ref{WDW-constr}), because of the noncommutativeness of
operators. The more consistent way to avoid the problems arising from the
noncommutativeness of operators is to impose the constraints
$\pi=0$ at the classical level before quantization. As was pointed out
in Sec.\,\ref{dynamics}, if the constraint $\pi=0$ holds, one comes from
(\ref{pi-Ham}) to the secondary constraint ${\cal T}=0$.
It is exactly what is implied in the BFV approach. Indeed, originally
the BFV approach was developed for constructing the S-matrix of an arbitrary
constrained system, \cite{BFV1, BFV2, BFV3, BFV4} i. e. the existence of
asymptotic states was supposed, therefore, the constraints
$\pi=0,\;{\cal T}=0$ have to hold.

There is still the third way to obtain the Wheeler -- DeWitt equation
(\ref{WDWeqn}) -- (\ref{R}) mentioned in the footnote \ref{zero-boundary},
p.~\pageref{zero-boundary}, Without adducing appropriate calculations let us
note that on this way two incorrect mathematical structures are necessarily
used: 1) a gauge-invariant path integral approximated on partially
degenerate action extremals without pointing out a procedure of removing
coordinate effects; 2) the definition of a wave function through a
divergent integral over gauge variables.

The derivation of the Wheeler -- DeWitt equation from a path integral was
investigated by Halliwell \cite{Hall}. There were many references to the
paper \cite{Hall}, so it seems to be relevant to comment it briefly.
The feature of the consideration presented in \cite{Hall} is that the
assumption about asymptotic states was made, so the Fradkin -- Vilkovisky
theorem was presumed to be valid for the case of a closed universe as well.
As a consequence of independence of the path integral on a gauge condition,
a particular gauge, $\dot N=0$, was chosen. At the same time, as it has
been demonstrated in the present paper, if the same calculations had
been made for an arbitrary gauge condition, it would have led to an equation
containing information about the chosen gauge.

The described methods of deriving the Wheeler -- DeWitt equation reveal the following.
\begin{enumerate}
\item The wave function not containing information about correlations between the
physical object and OM is the same in all the approaches.
\item In accordance with Sec.\,5, if the state vector
allows to compute average values of observable quantities, then information
about the correlations has to be contained in it inevitably. In other words,
there is no {\em physical} (normalizable) quantum state without a GVC.
\end{enumerate}

It is to be stated that there is no QGD as a gauge-invariant theory of
physical states in a closed universe based on the general principles of
QT. The illusion of the existence of such a theory arises if one forgets about
the necessity of singling out gauge orbit representatives and tries to come 
to an agreement about some special quantization rules. As to the correct path integral
method, it shows unambiguously that being applied to a closed universe, the 
ordinary QT of gravity is gauge-noninvariant. And this feature of the theory
is the evidence of its {\em adequacy} to the phenomena in question: it 
answers to the conditions of observations in a closed universe, where there is 
no possibility to remove an instrument for infinite distance from the object 
and thus to avoid influence of inertial fields, locally indistinguishable from 
a gravitational field, on the instrument. As it was mentioned in Sec.\,6,
a transition to another reference system (RS) implies a physical operation
in the whole Universe scale, and one hardly can think that such an operation 
could take place without physical consequences in view of nonvanishing
correlations between properties of the object and those of OM.

Some notions about the physical content of the QGD based on Eqs.\,(\ref{SE1}),
(\ref{H_full}), (\ref{V}) one can get from the presented in the next section
exact solution to this equation corresponding to the considered in
Sec.\,\ref{Class.exact} conditionally-classical solution.

\sect{The exact solution to the Schr\"odinger equation}
\label{SE_exact}
The task of constructing the wave function (\ref{time-depend.WF}) is reduced to
searching for a solution to the stationary equation (\ref{station.phys.SE})
for the physical part of the wave function under the parametrization-and-gauge condition
(\ref{simplegauge}). This equation reads
\begin{equation}
\label{SEphys}
-\frac12\,\frac{\partial^2\Psi_k}{\partial Q_a\partial Q^a}
 +U(Q^a)\Psi_k(Q^a)=E_k\Psi_k(Q^a),
\end{equation}
$Q^a=(q,\,\phi,\,\chi)$. Substitution (\ref{z1,z2}) enables to separate the 
variables in the equation, and it can be written in the following manner
$$
\left(6\hat{L}_1-\frac32\hat{L}_2+\frac12\hat{L}_3-kE\right)
 \Psi_k(z_1,\,z_2,\,\phi)=0,
$$
where
$$
\begin{array}{l}
\hat{L}_1=-\displaystyle\frac{\partial^2}{\partial z_1^2}
 +\frac19\exp(z_1),\\
\hat{L}_2=-\rule{0pt}{20pt}\displaystyle\frac{\partial^2}{\partial z_2^2}
 +\frac{16}9\exp(z_2),\\
\hat{L}_3=-\rule{0pt}{20pt}\displaystyle\frac{\partial^2}{\partial\phi^2}.
\end{array}
$$

The eigenfunctions of the operators $\hat{L}_1,\,\hat{L}_2$ appropriate to the 
positive eigenvalues 
$\displaystyle\frac{\nu^2_1}4,\,\displaystyle\frac{\nu^2_2}4$ 
are modified Bessel functions with an imaginary index,
$$
\psi_{\nu}(z)=\frac1{\sqrt{2\pi\,\Gamma(i\nu)}}K_{i\nu}
 \left[A\exp\left(\frac z2\right)\right];\quad
(A_1;\,A_2)=\left(\frac23;\,\frac83\right).
$$

So far as
$$
\exp\left(\frac{z_1}2\right)=4c^2;\quad
\exp\left(\frac{z_2}2\right)=4ac,
$$
the quantum number $\nu_1$ determines probability distribution for the scale 
$c$, and so does the quantum number $\nu_2$ for the scale $a=b$ at a given
$c$ value. Note, that for any $\nu$ there exist a quasiclassical solution 
to the problem,
$$
\mbox{\hbox to\textwidth{$
\psi_{\nu}(z)
 =\displaystyle\frac1{\sqrt{2}\,\Gamma(i\nu)\exp\left(\displaystyle
   \frac{\pi\nu}2\right)\left(\displaystyle\frac{\nu^2}4
   -\frac{A^2}4\exp(z)\right)^{\textstyle\frac14}}\times
$ \hfil}}
$$
$$
 \times\cos\left[2\left(\sqrt{\frac{\nu^2}4-\frac{A^2}4\exp(z)}
  -\frac{\nu}2\mathop{\rm Artanh}\sqrt{1-\frac{A^2}{\nu^2}\exp(z)}\right)
  +\frac{\pi}4\right],\quad
z<z_{\nu};
$$
$$
\mbox{\hbox to\textwidth{$
\psi_{\nu}(z)
 =\displaystyle\frac1{2\sqrt{2}\,\Gamma(i\nu)\exp\left(\displaystyle
   \frac{\pi\nu}2\right)\left(\displaystyle\frac{A^2}4\exp(z)
   -\frac{\nu^2}4\right)^{\textstyle\frac14}}\times
$ \hfil}}
$$
$$
 \times\exp\left[-2\left(\sqrt{\frac{A^2}4\exp(z)-\frac{\nu^2}4}
  -\frac{\nu}2\arctan\sqrt{\frac{A^2}{\nu^2}\exp(z)-1}\right)\right],\quad
z>z_{\nu},
$$
$z_{\nu}=\ln\left(\displaystyle\frac{\nu^2}{A^2}\right)$ 
being a classical turning-point.

The eigenfunctions of the operator $\hat{L}_3$ are plane waves
$$
\psi_p(\phi)=\frac1{\sqrt{2\pi}}\exp(ip\phi)
$$
that is in agreement with the classical solution (\ref{dotphi,Cs}).

The general solution to Eq.\,(\ref{SEphys}) for a given value of the parameter $E_k$,
describing the GVC state, is a superposition
\begin{equation}
\label{GSexact}
\Psi_{Ek}(z_1,\,z_2,\,\phi)=
 \int\!\!\!\!\int\limits_{-\infty}^{\infty}\!\!\!\!\int dp\,d\nu_1\,d\nu_2\,
  c_1(\nu_1,\,\nu_2,\,p)\,\psi_{\nu_1}(z_1)\,\psi_{\nu_2}(z_2)\,
  \psi_p(\phi)\,\delta\left(\frac32\nu^2_1-\frac38\nu^2_2
   +\frac12p^2-E_k\right).
\end{equation}

However, the stationary states which the wave functions (\ref{GSexact})
correspond to are not physical being unnormalizable since $E_k$ has a
continuous spectrum. A physical state is described by a time-dependent wave 
packet:
\begin{equation}
\label{exactE-pack}
\Psi_k(z_1,\,z_2,\,\phi,\,t)
 =\int\limits_{-\infty}^{\infty}dE_kc_2(E_k)\Psi_{Ek}(z_1,\,z_2,\,\phi)
  \exp\left[-iE_k(t-t_0)\right].
\end{equation}

Let us note, that in expressions (\ref{GSexact}), (\ref{exactE-pack}) the
quantity $E_k$ appears to be a controlling parameter providing, through the 
$\delta$-function, the correlation of the quantum numbers $\nu_1,\,\nu_2,\,p$, 
and, by that, a probability distribution of space scales at the quantum stage 
of the Universe evolution.

One can obtain the classical evolution law by computing the mean values of
the operators 
$\exp\left(\displaystyle\frac{z_1}2\right)$,\,
$\exp\left(\displaystyle\frac{z_2}2\right)$ 
over the packet (\ref{exactE-pack}). To do it the matrix elements will be 
required,
$$
\mbox{\hbox to\textwidth{$
\displaystyle\int\limits_{-\infty}^{\infty}dz\exp\left(\frac z2\right)
  \psi^*_{\mu}(z)\psi_{\nu}(z)
 =\left[2\pi\Gamma(-i\mu)\Gamma(i\nu)\right]^{-1}
  \int\limits_{-\infty}^{\infty}dz\exp\left(\frac z2\right)
   K_{-i\mu}\left[A\exp\left(\frac z2\right)\right]
   K_{i\nu}\left[A\exp\left(\frac z2\right)\right]=
$ \hfil}}
$$
\begin{equation}
\label{matr.elem}
 =\pi\left[4A\,\Gamma(-i\mu)\,\Gamma(i\nu)\,\right]^{-1}
  \left\{\cosh\left[\frac{\pi}2(\mu+\nu)\right]
   \cosh\left[\frac{\pi}2(\mu-\nu)\right]\right\}^{-1}.
\end{equation}

For the packet (\ref{GSexact}) -- (\ref{exactE-pack}) to describe really a
classically evolving Universe, it needs to be sufficiently narrow, i.e.
$c_1(\nu_1,\nu_2,p)$ and $c_2(E_k)$ do not have to deviate from zero values
beyond a small vicinity of their arguments near 
$(\bar{\nu}_1,\,\bar{\nu}_2,\,\bar{p})$ and $\bar{E}_k$. Therefore,
\begin{equation}
\label{mu,nu,omega}
\mu+\nu\approx2\bar{\nu};\quad
\mu-\nu\approx\frac{A\omega}{2\bar{\nu}},
\end{equation}
where $\omega=A^{-1}(\mu^2-\nu^2)$ is the difference between two values of 
the parameter $E_k$ corresponding to the quantum numbers $\mu$ and $\nu$. 
Note that the matrix element (\ref{matr.elem}) depends weakly on $\nu$ and 
decreases quickly when $|\mu-\nu|$ increasing. So, making use of the 
approximations (\ref{mu,nu,omega}) one obtains for the average of the
exponents $\exp\left(\displaystyle\frac z2\right)$,
\begin{equation}
\label{aver.exp}
\overline{\rule{0pt}{14pt}\exp\left(\frac z2\right)}
 =\frac14\tanh(\pi\bar{\nu})\int\limits_{-\infty}^{\infty}d\omega\,
  \frac{\exp\left[-i\omega(t-t_0)\right]}
   {\cosh\left(\displaystyle\frac{\pi A\omega}{4\bar{\nu}}\right)}
 =\frac{\bar{\nu}\tanh(\pi\bar{\nu})}
   {A\cosh\left[(2A^{-1}\bar{\nu}(t-t_0)\right]}.
\end{equation}
In the classical limit $\bar{\nu}$ is large, hence
$\tanh(\pi\bar{\nu})\approx1$, and, comparing (\ref{aver.exp}) with the 
classical expression (\ref{solutions}), one concludes that
$$
\alpha=\frac{\bar{\nu}_2}{A_2}=\frac38\bar{\nu}_2;\quad
\beta=\frac{\bar{\nu}_1}{A_1}=\frac32\bar{\nu}_1.
$$

From (\ref{GSexact}) the equation for the mean values follows $(k=0)$:
\begin{equation}
\label{Eaver}
\frac32\bar{\nu}_1^2-\frac38\bar{\nu}_2^2+\frac12\bar{p}^2=\bar{E},
\end{equation}
because $\overline{\nu_1^2}\approx\bar{\nu}_1^2$ an so on. The comparison of 
(\ref{alpha,beta-invar}) and (\ref{Eaver}) gives
$$
\bar{p}=C_s.
$$

Evidently, to every classical cosmological evolution scenario some 
configuration of the wave packet (\ref{GSexact}) -- (\ref{exactE-pack}) must
correspond, though not all solutions to the Schr\"odinger equation describe classical
universes, and, in addition, even those wave packets, for which the 
transition to the quasiclassical regime is possible, may turn to be unstable. 
Therefore in our approach the known problem of initial conditions for 
classical evolution is formulated as the problem of choice of quantum state
of the Universe. The quantum state in the Bianchi-IX model is determined by a 
concrete kind of the function
$$
\tilde{C}(\nu_1,\,\nu_2,\,p,\,E_k)=c_2(E_k)\,c_1(\nu_1,\,\nu_2,\,p),
$$
describing a wave packet structure.

We do not know how the choice of the quantum state is made.
Perhaps, it is made according to statistical laws in the process of the
creation of the Universe from ``Nothing". In Sec.\,\ref{Creation} we will
discuss the hypothesis according to which the act of the creation is thought
as a quantum transition taking place out of time from the special singular 
state to one of the physical states of the Universe which the wave packets 
(\ref{GSexact}) -- (\ref{exactE-pack}) correspond to.

A quasiclassical wave packet may also be written in the form
$$
\mbox{\hbox to\textwidth{$
\displaystyle\Psi_k(z_1,\,z_2,\,\phi,\,t)
 =\int\!\!\!\!\int\limits_{-\infty}^{\infty}\!\!\!\!\int dE_kdp\,d\nu_1d\nu_2
  \tilde{C}(\nu_1,\,\nu_2,\,p,\,E_k)
  \exp\left[-iE_k(t-t_0)+i\sigma_1(z_1)+i\sigma_2(z_2)\right]\times
$ \hfil}}
$$
$$
  \times\psi_p(\phi)\,
  \delta\left(\frac32\nu_1^2-\frac38\nu_2^2+\frac12p^2-E_k\right).
$$
Here the sum $\sigma_1(z_1)+\sigma_2(z_2)$ is the part of the classical action
$S(z_1,\,z_2,\,\phi,\,t)$, determining its dependence on $z_1$ and $z_2$ (the 
scalar field is treated as essentially quantum). The functions $\sigma(z)$ 
satisfy the equations
\begin{equation}
\label{sigma}
\frac{\partial\sigma}{\partial z}
 =\sqrt{\frac{\bar{\nu}^2}4-\frac{A^2}4\exp(z)}.
\end{equation}

If the dependence of the classical action on the variables $z$ is given one
can reconstruct the evolution law (\ref{solutions}) with the help of the 
standard procedure. But the two mentioned methods of going over to the 
classical limit are applicable owing to the explicit dependence of the general solution on
time. And this, in turn, is due to the indication of the concrete choice of 
a RS where the time $t$ is measured, available in the theory.

In the classical limit a classical subsystem of the physical object itself can
be considered as a RS. Such a subsystem cannot fill the whole space; it is
admissible that it occupies a limited region of space. So, we will refer
to such RS as to a local one. The appearance of the time $\tau$ introduced
as a parameter along a classical path is associated with this very RS.

A derivative with respect to path length can be defined by the following way:
$$
\frac d{dt}=u(\tau)\nabla S\,\nabla,
$$
where $u(\tau)$ is an arbitrary function, $\nabla S$ is a tangent vector to 
the path;
$$
\nabla=\left(2\sqrt{3}\,\frac{\partial}{\partial z_1},\quad
 i\sqrt{3}\,\frac{\partial}{\partial z_2}\right).
$$
On the other hand,
$$
\frac d{d\tau}=\frac{d z_1}{d\tau}\frac{\partial}{\partial z_1}
 +\frac{d z_2}{d\tau}\frac{\partial}{\partial z_2},
$$
whence
$$
\frac{d z_1}{d\tau}=12u(\tau)\frac{d\sigma_1}{d z_1};\quad
\frac{d z_2}{d\tau}=-3u(\tau)\frac{d\sigma_2}{d z_2}.
$$
From (\ref{sigma}) one obtains
$$
\frac{d z_1}{d\tau}=4u(\tau)\sqrt{\beta^2-\exp(z_1)};\quad
\frac{d z_2}{d\tau}=-4u(\tau)\sqrt{\alpha^2-\exp(z_2)}.
$$
and, as a result of integration,
\begin{equation}
\label{tau-solut}
\exp\left(\frac{z_1}2\right)
 =\beta\cosh^{-1}\left(2\beta\left[\tilde{u}(\tau)-\tau'_0\right]\right);\quad
\exp\left(\frac{z_2}2\right)
 =\alpha\cosh^{-1}\left(2\alpha\left[\tilde{u}(\tau)-\tau_0\right]\right);
\end{equation}
$$
\tilde{u}(\tau)=\int u(\tau)d\tau;\quad
\tau_0,\,\tau'_0={\rm const}.
$$

The time $\tau$ of a local observer emerges irrespectively of the existence 
of the time $t$, but both the times may correlate between each other. To 
bring in correspondence the expressions (\ref{solutions}) and 
(\ref{tau-solut}) it is sufficient to put $u(\tau)=1$.

In ordinary quantum mechanics the time involving in the Schr\"odinger equation is also associated
with an observer making measurements on a physical system. That time may
fail to coincide with the time appearing in Heisenberg operator equations, as
well as with the world time in which dynamics of the system can be described
in case the system is quasiclassical. In quantum mechanics the hypothesis
about the equivalence of the mentioned times is used, though this is nowhere
specified. In the considered example we have manifested that the times used
for describing the evolution of a quantum system are different in general. In
QGD times associated with different observers can be brought to agreement
with each other by choice of a gauge condition.

\sect{The problem of the creation of the Universe from ``Nothing"}
\label{Creation}
In the previous section an EPS scheme has been proposed that enables one to
describe phenomenologically quantum evolution of the Universe as an
integrated system including OM. But we have not touched upon the main
question of quantum cosmology -- the problem of the creation of the Universe
from ``Nothing" -- and the associated problem of finding the initial state
for cosmological evolution.

There are several approaches to the question about the quantum origin of the
Universe, among the most significant works the papers by Hartle and Hawking
\cite{Hartle} and Vilenkin \cite{Vil1,Vil2,Vil3} should be mentioned. These
works are based on the Wheeler -- DeWitt equation in the minisuperspace of
two dimensions -- the scale factor of the Friedman -- Robertson -- Walker
space and a homogeneous scalar field. The creation of the Universe is
thought to be a quantum tunneling from a classically forbidden region of
the minisuperspace, and the state ``Nothing" is described by the asymptotic
of a wave function corresponding to the vicinity of scale factor values of
Planckian order. The discussion was concentrated on the choice of boundary
conditions for a wave function of the Universe, namely, the ``no boundary''
and tunneling prescription for the wave function.

The interest to the problem of creation of the Universe was aroused
two years ago by the proposal of Hawking and Turok that the
``no boundary'' wave function may describe creation of an open inflationary
universe \cite{HT1}. The attempts were done to make cosmological predictions
on the base of this proposal \cite{HT2, HR}. At the same time, Hawking --
Turok approach was seriously criticized by Linde \cite{Linde, HT3, BL} and
Vilenkin \cite{Vil4} (see especially Linde's review on the problem of
quantum creation of the Universe, \cite{Linde}).

In the other approach \cite{McGu,Hos} the creation of the Universe (more
precisely, universes) is considered from the position of the second
quantization (the third, regarding matter fields). The initial state is a
vacuum of universes, and the creation act is analogous to particle creation in
a nonstationary metric, the superspace metric playing its role. The 4-metric
determinant  (in the minisuperspace -- the scale factor) acts as a time
parameter causing nonstationarity of the metric. But the rightfulness of
identification of this parameter with time beyond a quasiclassical region
arouses doubt.

On the other hand, according to Grishchuk and Zeldovich \cite{Grish}, in the
initial state, the state ``Nothing", there is neither space with its
geometry nor time. The creation of the Universe is essentially a
quantum-gravitational process in the result of which a classical space-time
comes into existence. Hence, the process cannot be ordered by a geometric
parameter, as it is done, for instance, in the approach based on the third
quantization. There is no reason to employ some physical parameters to
define the state ``Nothing" as well.

Thus, the state ``Nothing" belongs to a deeper level of reality beyond the
bounds of the physical reality which quantum mechanics was formulated 
for. Strictly speaking, we have no grounds to describe such a state by a
wave function obeying an Schr\"odinger equation.

But we cannot avoid making an attempt to apply quantum-mechanical concepts 
for describing the creation of the physical reality itself%
\footnote{Similarly, it is impossible to reject classical notions (among
which we live) such as, for instance, particle coordinates and momenta,
when interpreting quantum mechanics.}.

It may be supposed that the Universe appears as a result of spontaneous
reduction of the state ``Nothing", potentially containing all the possible 
Universe states, to one of them. (Such a picture, as a matter of fact, is 
present in the Penrose's report \cite{Pen}). The wave function of the state ``Nothing"
is a superposition of all physical states but is not normalizable itself.
From the classical viewpoint, the initial state of the Universe is a singular 
state without structure. From the QGD viewpoint, the Universe is localized in
the region of small values of scale factor in minisuperspace. So far as this
state cannot have any internal structure, there is no gravitation wave or
matter field in it. Evidently, there exist no observer either. The wave function of the
state must satisfy an equation describing an object with the only degree of
freedom -- the scale factor $r=\exp(q/2)$. This equation cannot be derived on 
the grounds of general theoretical principles since all of them anyway appeal 
to space-time notions. It can be only postulated on the basis of the above 
ideas.

We postulate the equation for the wave function of the state ``Nothing" as follows:
\begin{equation}
\label{Nothing-eqn}
\frac{d^2\Psi_0}{dq^2}-4\exp(2q)\Psi_0=0.
\end{equation}
It can be ``obtained" from Eq.\,(\ref{SEphys}) by excluding all degrees of
freedom but the scale factor. Its form, however, must not depend on a
chosen concrete model. The similar equation for the wave function of the
state ``Nothing'' was firstly suggested by Vilenkin \cite{Vil1}.

The solution to Eq.\,(\ref{Nothing-eqn}) is the modified Bessel function of
the zero order
\begin{equation}
\label{Noth.WF}
\Psi_0(q)=C_0K_0\left[2\exp(q)\right],
\end{equation}
possessing all the mentioned properties; it describes a state with no matter,
no observer and, consequently, no time, and with no space itself: the Universe
is ``locked" in the isotropic singularity $q=-\infty\;(r=0)$.

Relative probabilities of transitions from the state ``Nothing" to physical 
states, which the wave packets (\ref{GSexact}), (\ref{exactE-pack})
correspond to, at $t=t_0$ are given by the projections of the packets on 
the wave function (\ref{Noth.WF}):
$$
\int dq\,d\chi\,d\phi\,d\mu\,\Psi_0^*(q)
 \int dk\Psi_k(q,\,\chi,\,\phi,\,t=t_0)\,\delta(\mu-k)
 =\int dE\,d\nu_1d\nu_2\,E^{-1}\sqrt{2\pi}\,
  \tilde{C}(\nu_1,\nu_2,0,E)\,\xi(\nu_1,\nu_2),
$$
$$
\xi(\nu_1,\nu_2)
 =\int dq\,d\chi\,\Psi_0^*(q)\,\psi_{\nu_1}(q,\chi)\,\psi_{\nu_2}(q,\chi).
$$

The time moment $t=t_0$ corresponds to the creation of the Universe, the act 
of the quantum transition from the state ``Nothing" to one of the physical
states itself taking place out of time. The act results in the emergence of
an observer (the GVC). From this moment we can consider the evolution of the
Universe in time, described in our model by changing in time of the wave 
packet (\ref{GSexact}), (\ref{exactE-pack}), with the supposed subsequent
going over to a quasiclassical regime.

Thus, the problem of the creation of the Universe from ``Nothing" becomes a
computational problem of transition amplitudes to various physical states.

\sect{Conclusions}
\label{Conclus}
Let us briefly formulate the main results of the work.
\begin{enumerate}
\item It is shown that the standard Wheeler -- DeWitt QGD fails to be constructed by correct
mathematical methods of quantum theory based on the path integral formalism.
The main physical cause for the Wheeler -- DeWitt QGD nonexistence in the frames of general
QT is that there are no asymptotical quantum states providing non-interference 
of measurement means in interaction processes of gravitational fields.
\item A mathematically correct QGD of the Bianchi-IX model is formulated
in extended phase space (EPS); the basic equation of the QGD in EPS is the 
dynamical Schr\"odinger equation describing time evolution of a physical
object (gravitational fields of a closed universe) and a classical 
gravitational vacuum condensate (GVC) breaking symmetry of the system after 
introducing continually distributed observation means.
\item Gauge noninvariance is inherent in the proposed version of QGD in EPS in
a fundamental manner; in the frames of the Copenhagen operational
interpretation of QT it is shown that this property {\em adequately} reflects
the conditions of observations in a closed universe in mental experiments
operating with the notions associated with the Landau-Lifshitz reference systems.
\item On the example of the exactly solvable quantum Bianchi-IX model with 
one gravitational-wave degree of freedom it is shown that the GVC determines a 
cosmological scenario character through the controlling parameter -- its whole 
energy.
\item The new approach to the creation of the Universe as an objective
reduction of the singular state ``Nothing" to one of alternative physical
states existing in the gauge-noninvariant QGD in EPS is proposed.  It is shown
that in this version there is the possibility to evaluate numerically relative 
probabilities of various physical (normalizable) initial states (at the
moment $t=t_0$) of quantum cosmological evolution.
\end{enumerate}

The proposed QGD version we consider as a phenomenological extrapolation of 
the existing QT methodological principles and formalism on scales
of a closed universe as a whole. Evidently, such an extrapolation, even being
mathematically correct, cannot be physically complete. The phenomenologicness
of QT itself, the lack of an answer to the question about nature of quantum
integrity of a physical object and OM inside it shows the necessity to make
choice of mathematical procedures, finally defining the formalism of the
theory, without clear understanding the motives for this choice. In 
particular, the number of unsettled problems includes:
\begin{enumerate}
\item The operator ordering problem, or the mathematically equivalent to it 
problem of the choice of path integral approximation.
\item The problem of uncertainty about the choice of a parametrization fixing
gauge variables, and of gauge conditions realizing a choice of a reference
system in the Landau-Lifshitz RS class; physically, this problem
consists in the lack of strict principles fixing the GVC properties.
\end{enumerate}

The existence of these problems let the question about quantitative
characteristics of quantum correlations between the properties of a physical
object and those of OM be open; the common origin of the problems is
insufficient understanding what the measuring process in quantum gravity is.
To solve the problems a self-consistent theory is required that primordially
gives a description of the integrated system including OM, and contains a
quantum theory of measurements as a constituent part.

We suppose that in the future theory a solution to the mentioned above
problems will be obtained on the grounds of unification of the geometric and
the quantum integrity conceptions. As to the existing phenomenological QT
that is in principle gauge-noninvariant for a closed universe, we are forced
to bind it to a concrete gauge class, and to bind the procedure of solving
equations to a concrete gauge. In this situation for making gauge choice it is
necessary to have some argumentation, ours consisting in the following:
\begin{enumerate}
\renewcommand{\labelenumi}{\theenumi)}
\item the possibility of comparing with the canonical operator formalism;
\item the possibility to interpret directly the Landau-Lifshitz RS;
\item the availability of the wave properties of the GVC excitations.
\end{enumerate}

The latter fixes a {\em representative} of a gauge class. From the future
theory, likely, a deeper criteria for gauge choosing should be expected .

Let us note, finally, that after Penrose \cite{Pen,Penr} having discussed the
necessity for a radical change of the QT content, we suppose that the new 
theory will require essentially new ideas and mathematical algorithms. 
According to its status, the theory should include a description of
irreversible measurement processes; therefore, it should have one more
substantial feature -- {\em time irreversibility}. In other words, the
presence of such a subsystem as a GVC in an integrated system must, likely,
{\em break Hermitianness} of Hamiltonian of the system.

\end{document}